\documentclass[12pt]{article}
\usepackage{fullpage,url}
  \usepackage{amssymb}
	\usepackage{graphicx}

\newtheorem{theorem}{Theorem}

\def\diag{\mathrm{diag}}
 
\def\arccosh{\mathrm{arccosh}}

\def\st{:}
\def\ball{\mathrm{ball}}

\def\calP{\mathcal{P}}

\def\calK{\mathcal{K}}

\def\bbR{\mathbb{R}}
\def\bbB{\mathbb{B}}

\def\bbL{\mathbb{L}}

\def\innerL#1#2{{\langle #1,#2\rangle}_L}
\def\inner#1#2{{\langle #1,#2\rangle}}

\def\innerL#1#2{{\langle #1,#2\rangle}_\mathrm{L}}

\def\Vor{\mathrm{Vor}}

\def\path#1{}

\def\bbB{\mathbb{B}}
\def\myvec#1#2{ \left[\begin{array}{c}#1\cr #2\end{array}\right]}
 \def\calF{\mathcal{F}}
 
\begin{document}

\title{Further results on the hyperbolic Voronoi diagrams}

\date{April 2014} 

\author{Frank Nielsen\thanks{Ecole Polytechnique, France 
Sony Computer Science Laboratories, Japan.
Email: {\tt Frank.Nielsen@acm.org}}
\and
Richard Nock\thanks{NICTA, Australia.
UAG CEREGMIA, France.
Email: {\tt rnock@martinique.univ-ag.fr}}
}

\maketitle

\begin{abstract}
In Euclidean geometry, it is well-known that the $k$-order Voronoi diagram in $\bbR^d$ can be computed from the vertical projection of the $k$-level of an arrangement of hyperplanes tangent to a convex potential function in $\bbR^{d+1}$: the paraboloid.
Similarly, we report  for the Klein ball model of hyperbolic geometry such a {\em concave} potential function: the northern hemisphere.
Furthermore, we also show how to build the hyperbolic $k$-order diagrams as  equivalent clipped power diagrams in $\bbR^d$.
We investigate the hyperbolic Voronoi diagram in the hyperboloid model and show how it reduces to a Klein-type model using central projections.
\end{abstract}


\section{Introduction}
Hyperbolic geometry is a consistent geometry where the Euclidean Playfair's parallel postulate is discarded and replaced by the existence of many lines $U$ not intersecting another given line $L$  and passing through a given point $P\not\in L$ (the $U$'s are said {\em ultra-parallel}\footnote{{\em Parallel} lines intersect at infinity in hyperbolic geometry.} to $L$).
Hyperbolic geometry can be studied using various models~\cite{VHVD-2014}: Poincar\'e disk or upper plane conformal models, Klein  non-conformal model disk model, hyperboloid conformal model, etc.
From the viewpoint of computational geometry, we prefer to use Klein model where lines/bisectors are Euclidean straight~\cite{HVDeasy-2010} and then convert the output to the desired model for visualization or navigation purposes~\cite{VHVD-2014}. 
We report further novel results for constructing hyperbolic Voronoi diagrams (HVDs) in Klein model~\cite{HVDeasy-2010} and present yet another approach to get Klein-type affine bisectors/diagrams from the hyperboloid\footnote{\underline{Hyperbol}ic geometry stems from the \underline{hyperbol}oid model.} model.

%%%%%%%%
\section{HVDs from lower envelopes}
%%%%%%%%

The {\em Voronoi diagram} of  a set $\calP=\{p_1, ..., p_n\}$ of $n$ points  in $\mathbb{R}^d$ w.r.t. $D(\cdot,\cdot)$ can be computed equivalently as the {\em minimization diagram} of $n$ functions by observing that
$D(x,p_i)\leq D(x,p_j) \Leftrightarrow F_i(x)\leq F_j(x)$ where $F_l(x)=D(x,p_l)$, $l\in\{1, ..., n\}$.
Thus the {\em combinatorial structures} are congruent: $\Vor_D(\calP) \cong \min_{l\in\{1, ...,n\}} F_l(x)$.
Furthermore, this minimization diagram amounts to compute the {\em lower envelope} of $n$ graph functions in $\bbR^{d+1}$:
 $\mathcal{F}_l:\{(x,y=F_l(x))\ :\ x\in \mathbb{R}^d\}$.

Let $\inner{x}{p}=x^\top p=\sum_{i=1}^d x^{(i)} p^{(i)}$ denotes the Euclidean inner product.
In the Klein model~\cite{HVDeasy-2010}, the distance between two points $x$ and $p$ in the open unit ball domain $\bbB_d=\{ x\in\bbR^d \st \inner{x}{x}<1\}$  is $D^K(x,p)=\arccosh \frac{1-\inner{x}{p}}{\sqrt{1-\inner{x}{x}}\sqrt{1-\inner{p}{p}}}$ where
$\arccosh(x)=\log(x+\sqrt{x^2-1})$ for $x\geq 1$ is a monotonically increasing function.
Since the Voronoi diagram does not change by composing the distance with a monotonous function, we consider the equivalent Klein distance $d^K(x,p)=\frac{1-\inner{x}{p}}{\sqrt{1-\inner{x}{x}}\sqrt{1-\inner{p}{p}}}$.
To each point $p_i\in\calP$ corresponds a function $F_i(x)=d^K(x,p_i)$.
Since the denominator $\sqrt{1-\inner{x}{x}}$ is common to all functions, the minimization diagram is equivalent to the minimization diagram
of $F_i'(x)=\frac{1-\inner{x}{p_i}}{\sqrt{1-\inner{p_i}{p_i}}}$.
The graph $\calF_i'=\{ (x,y=F_i(x)) : x\in\bbB_d \}$ are {\em hyperplanes} in $\mathbb{R}^{d+1}$ defined on $\bbB_d$, and the lower envelope can thus be computed from the intersection of $n$ halfspaces $H_i^-: y\leq \frac{1-\inner{x}{p_i}}{\sqrt{1-\inner{p_i}{p_i}}}$, yielding the Voronoi unbounded polytope in $\bbR^{d+1}$.

\begin{theorem}
The HVD of $n$ points can be computed in the Klein model as the intersection of $n$ half-spaces in $\bbR^{d+1}$ and by projecting vertically ($\downarrow$ $H_0: y=0$) the polytope on $\bbR^d$, and clipping it with the unit ball domain: $\Vor_{d^K}(\calP)=((\cap_{i=1}^n H_i^{-})\downarrow H_0)\cap \bbB_d$.
\end{theorem}

%%%%%%%%
\section{Lifting sites to a potential function}
%%%%%%%%

In Euclidean (and more generally Bregman geometry), the Voronoi polytope is built by lifting points to tangent hyperplanes to a {\em potential function} $y=F(x)$ at site locations. This is the paraboloid lifting transformation:  $y=F(x)=\inner{x}{x}$ ($y=F(x)$ for a convex Bregman generator $F$).

\begin{theorem}
In the Klein ball model, the {\em potential function} for lifting generators to hyperplanes is the {\em concave} function 
$y=F(x)=\sqrt{1-\inner{x}{x}}$ restricted to $\bbB_d$.
\end{theorem}

Proof:
Let us identify the  hyperplane equation $H(p): y=\frac{1-\inner{p}{x}}{\sqrt{1-\inner{p}{p}}}$ with
the hyperplane tangent at $p$ to a potential function $y=F(x)$: 
$\inner{\nabla F(p)}{x-p}+F(p)=\inner{x}{\nabla F(p)} + F(p)-\inner{p}{\nabla F(p)}$.
We have $\nabla F(p)=-\frac{p}{\sqrt{1-\inner{p}{p}}}$ and the remaining term (independent of $x$) is $\frac{1}{\sqrt{1-\inner{p}{p}}}$.
The anti-derivative of $\nabla F(x)=-\frac{x}{\sqrt{1-\inner{x}{x}}}$ is $\sqrt{1-\inner{x}{x}}+c$, and the constant $c$ solves to zero.
This is the equation $y^2+\inner{x}{x}=1$ of the northern hemisphere for $y\geq 0$.
Observe that the hyperplanes tend to become vertical as we near the boundary domain $\partial\bbB_d$, and are vertical at the boundary.
 
\section{$k$-order hyperbolic Voronoi diagrams}

Since the Klein bisector is affine, the $k$-order HVD is affine. We present two construction methods.

\subsection{$k$-HVDs from levels of an  arrangement of hyperplanes}
This is a straightforward generalization of the Euclidean procedure using the $\sqrt{1-\inner{x}{x}}$ potential function.
The $k$-order HVD is a {\em cell complex} that can be built by projecting to $\mathbb{R}^d$ all the $(d+1)$-dimensional cells at $k$-level of the arrangement of the site hyperplanes $\mathcal{H}:\{H_1, ..., H_n\}$ of $\mathbb{R}^{d+1}$  and clipping the structure to $\bbB_d$.
Figure~\ref{fig:example} displays some $k$-order diagrams and illustrates some degenerate cases.

\def\ttt{0.33\columnwidth}
\begin{figure}%
\centering
\begin{tabular}{cc}
\includegraphics[bb=0 0 512 512,width=\ttt]{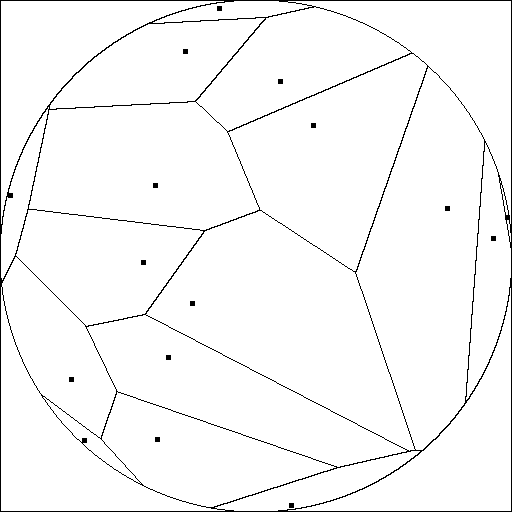} &
\includegraphics[bb=0 0 512 512,width=\ttt]{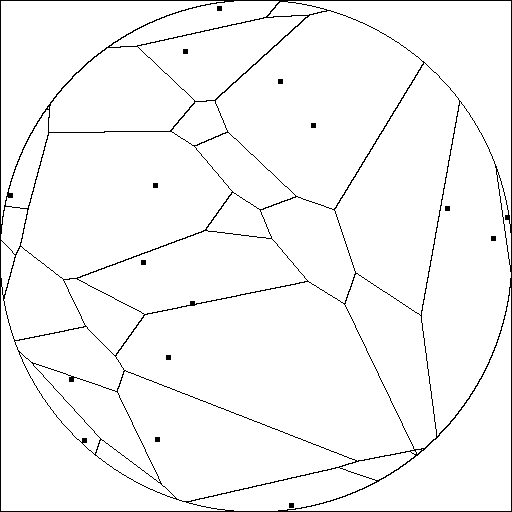} \\
(a) & (b) \\
\includegraphics[bb=0 0 1024 1024,width=\ttt]{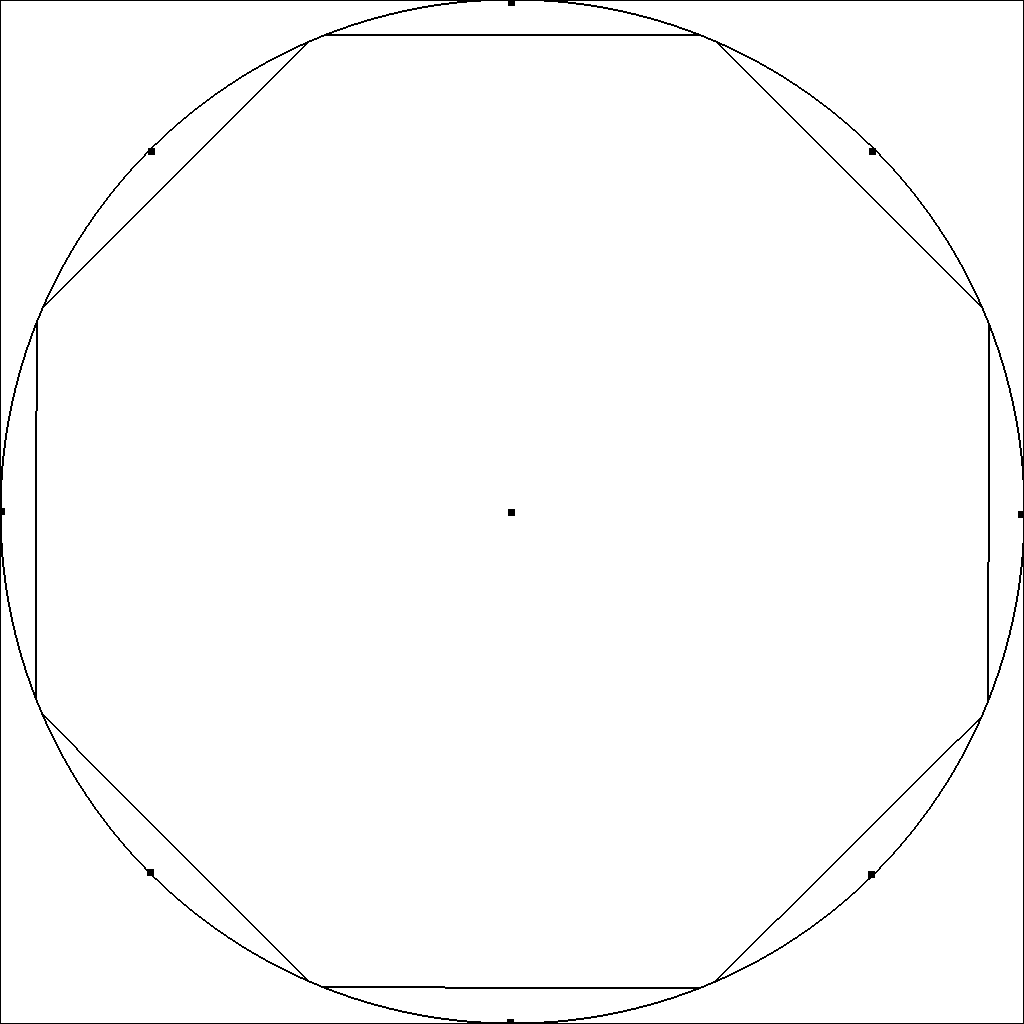} &
\includegraphics[bb=0 0 1024 1024,width=\ttt]{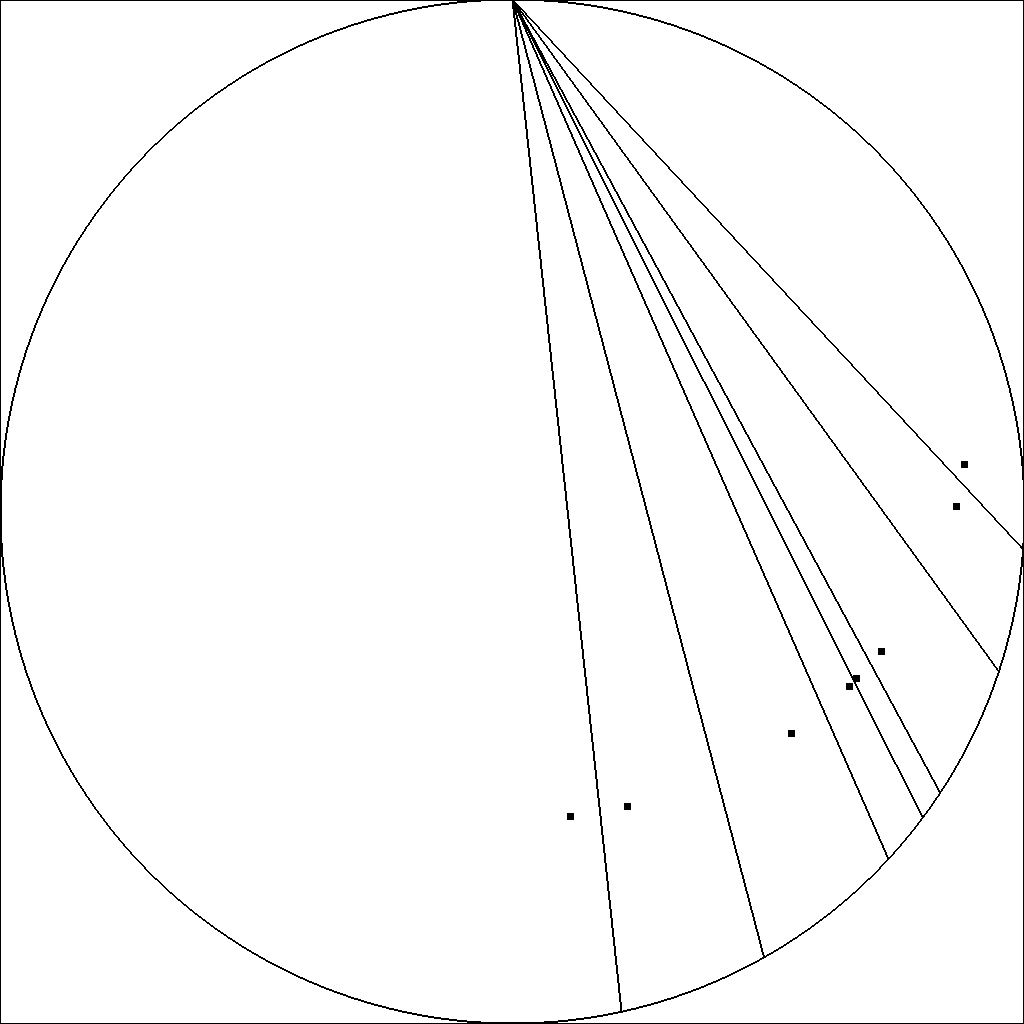} \\
(c) & (d)
\end{tabular}
\caption{HVD for $k=1$ (a) and $k=2$ (b).
HVD with all unbounded cells (c), and pencil of parallel bisectors intersecting at $\partial\bbB_d$ (d).
}%
\label{fig:example}%
\end{figure}
%Note that in the HVD, all cells may be unbounded as depicted in .
%This property is used for embedding trees, interpreted as dual Delaunay simplicial complexes, in hyperbolic geometry.
 
\subsection{$k$-HVDs from power diagrams}

Consider all subsets of size $k$, $\calP_k={\calP\choose k}=\{\calK_1, ..., \calK_N\}$ with $N={n\choose k}$.
Those {\em subset generators} partition the space into {\em non-empty $k$-order Voronoi cells}: 
$$
\Vor_k(\calK_i) = \{x\st \forall q\in\calK_i, \forall r\in \calP\backslash\calK_i,\ D(x,q)\leq D(x,r)\}.
$$

Observe that $x\in\Vor_k(\calK_i)$ iff $\sum_{p\in\calK_i} D(x,p) \leq \sum_{p'\in\calK_j} D(x,p')$.
In Klein model with $D=d^K$, we define the function $\sigma_{\calK_i}(x) = \sum_{x\in \calK_i} \frac{1-\inner{x}{p_i}}{\sqrt{1-\inner{p_i}{p_i}}}$, and
$x\in\Vor_k(\calK_i) \Leftrightarrow h_{\calK_i}(x)\leq h_{\calK_j}(x)\ \forall j\not=i$.
By identifying those hyperplane equations with the generic power diagram hyperplane $h(x): y=-2\inner{x}{c}-w+\inner{c}{c}$ for a ball centered at $c$ and radius $r^2=w$ ($r$ may be imaginary when $w<0$), we transform each $k$-subset $\calK_i$ in Klein model into a weighted point (or ball) $\ball(c_i,w_i)$:
$c_i=\sum_{p\in\calK_i} \frac{p}{2\sqrt{1-\inner{p}{p}}}$ and $w_i=\inner{c_i}{c_i}- \sum_{p\in\calK_i} \frac{1}{\sqrt{1-\inner{p}{p}}}$.
This method is only practical if when we consider all subsets $\calK_i$ that yields non-empty cells, otherwise we have $N={n\choose k}$ too many balls to be tractable!

%%%%%
\section{HVDs from the hyperboloid model}
%%%%%
Consider the symmetric bilinear form $L=\diag(-1,1, ...,1)$ in Minkowski space $\bbR^{1,d}$: $\inner{p}{q}_L=p^\top L q=-p^{(0)}q^{(0)}+\sum_{i=1}^d p^{(i)}q^{(i)}$. The hyperboloid model is defined on the upper sheet domain $\bbL^+=\{ \innerL{x}{x}=-1,\ x_0>0 \}$ (interpreted as a sphere $\innerL{x}{x}=R^2$ of imaginary radius $R=i$).
For $x\in\bbR^d$, we denote $x^L$ its point obtained by vertically rising $(\cdot,x)$ on $\bbL^+$: $x^L=(\sqrt{1+\inner{x}{x}},x)$, called Weierstrass coordinates. 
The hyperbolic distance is expressed by $D^L(p^L,q^L)=\arccosh (-\innerL{p^L}{q^L})$ and is equivalent to $d^L(p^L,q^L)=-\innerL{p^L}{q^L}$.
For two points $p^L$ and $q^L$ on $\bbL^+$, the bisector equation is $\innerL{x^L}{p^L-q^L}=0$.
The bisector is an hyperbola of equation $\left(\sqrt{1+\inner{p}{p}}-\sqrt{1+\inner{q}{q}}\right) \sqrt{1+\inner{x}{x}} + \inner{q-p}{x} = 0, x\in\bbR^d \ (*)$.
This hyperbola bisector is contained in a hyperplane $H(p,q)$ of $\bbR^{d+1}$ passing through the origin $O$:
$H(p,q) :  (\sqrt{1+\inner{p}{p}}-\sqrt{1+\inner{q}{q}}) x_0 + \inner{q-p}{x} = 0$.
The Klein disk model is obtained from $\bbL^+$ by a central projection $\pi$ from the origin to the hyperplane $H_1: x_0=1$: $\pi \myvec{x_0}{x}=\myvec{1}{x'=\frac{x}{x_0}=\frac{x}{\sqrt{1+\inner{x}{x}}}}$. The disk center touches the apex of $\bbL^+$.
Let $a_{p,q}=\sqrt{1+\inner{p}{p}}-\sqrt{1+\inner{q}{q}}$.
Multiplying $(*)$ by $\frac{1}{\sqrt{1+\inner{x}{x}}}$, we have the bisector written as
$\inner{q-p}{x'}+a_{p,q}=0$, an affine bisector in $x'$.

Now consider $\pi_{c,l}$ the {\em generic} central projection of $\bbL^+$ from $C=(c,0)$ to the hyperplane $H_l:x_0=l$ so that $\pi=\pi_{0,1}$.
We have
$\pi_c \myvec{\sqrt{1+\inner{x}{x}}}{x}  = \myvec{l}{x_{c,l}=\frac{l-c}{\sqrt{1+\inner{x}{x}}-c} x}, c\not = 1$.
Choosing $c=0$ and $0<l\leq 1$ yields the same construction procedure but the clipping of the equivalent power diagram~\cite{HVDeasy-2010} need to be done on a disk of size $l$ since
$\|x_{c,l}\|= \| \frac{l}{\sqrt{1+\inner{x}{x}}}x\| \leq l$, $\forall x\in\bbR^d$.

\begin{figure}%
\centering
\begin{tabular}{ccc}
\includegraphics[width=0.3\textwidth]{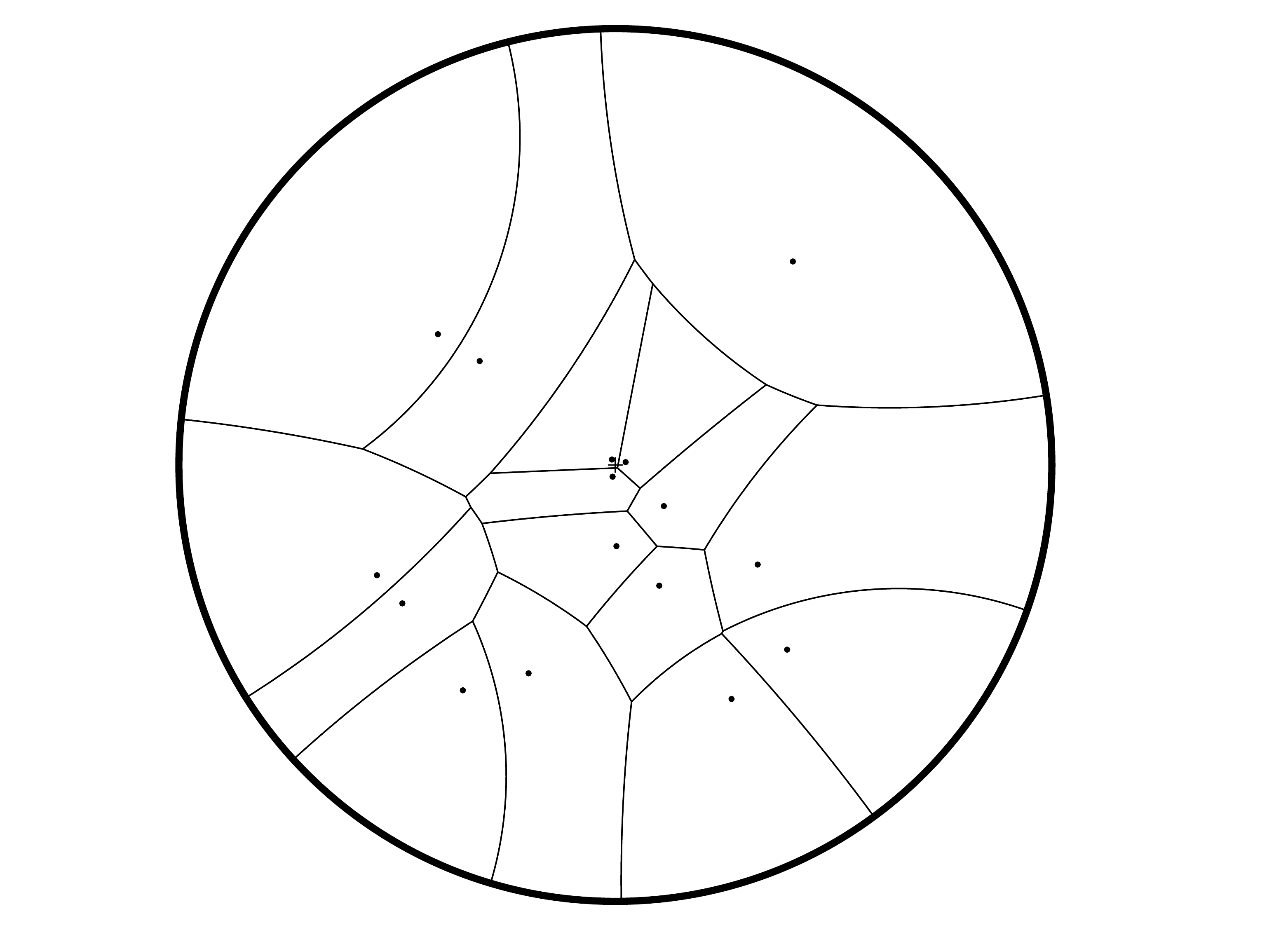} &
\includegraphics[width=0.3\textwidth]{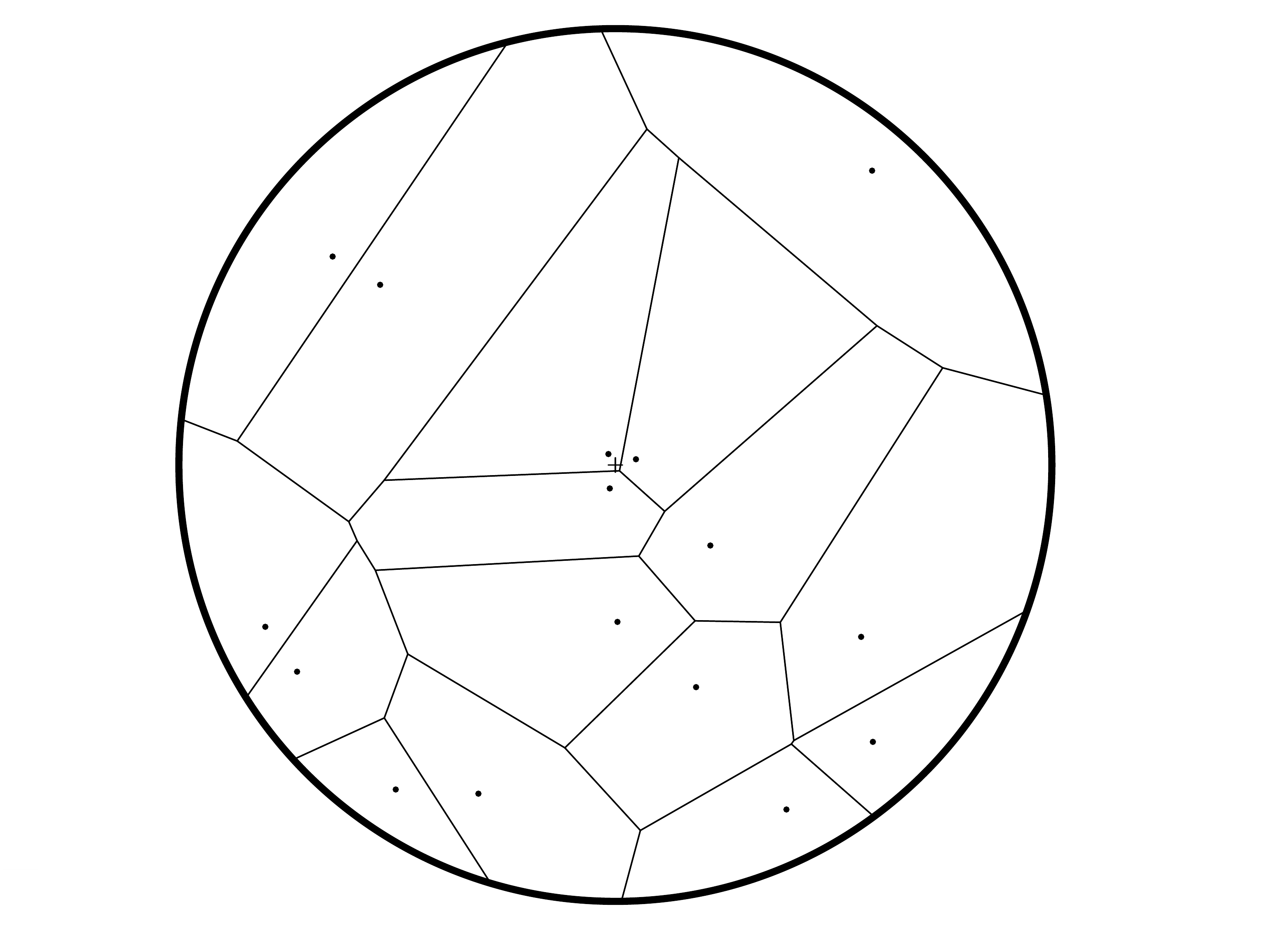} &
\includegraphics[width=0.3\textwidth]{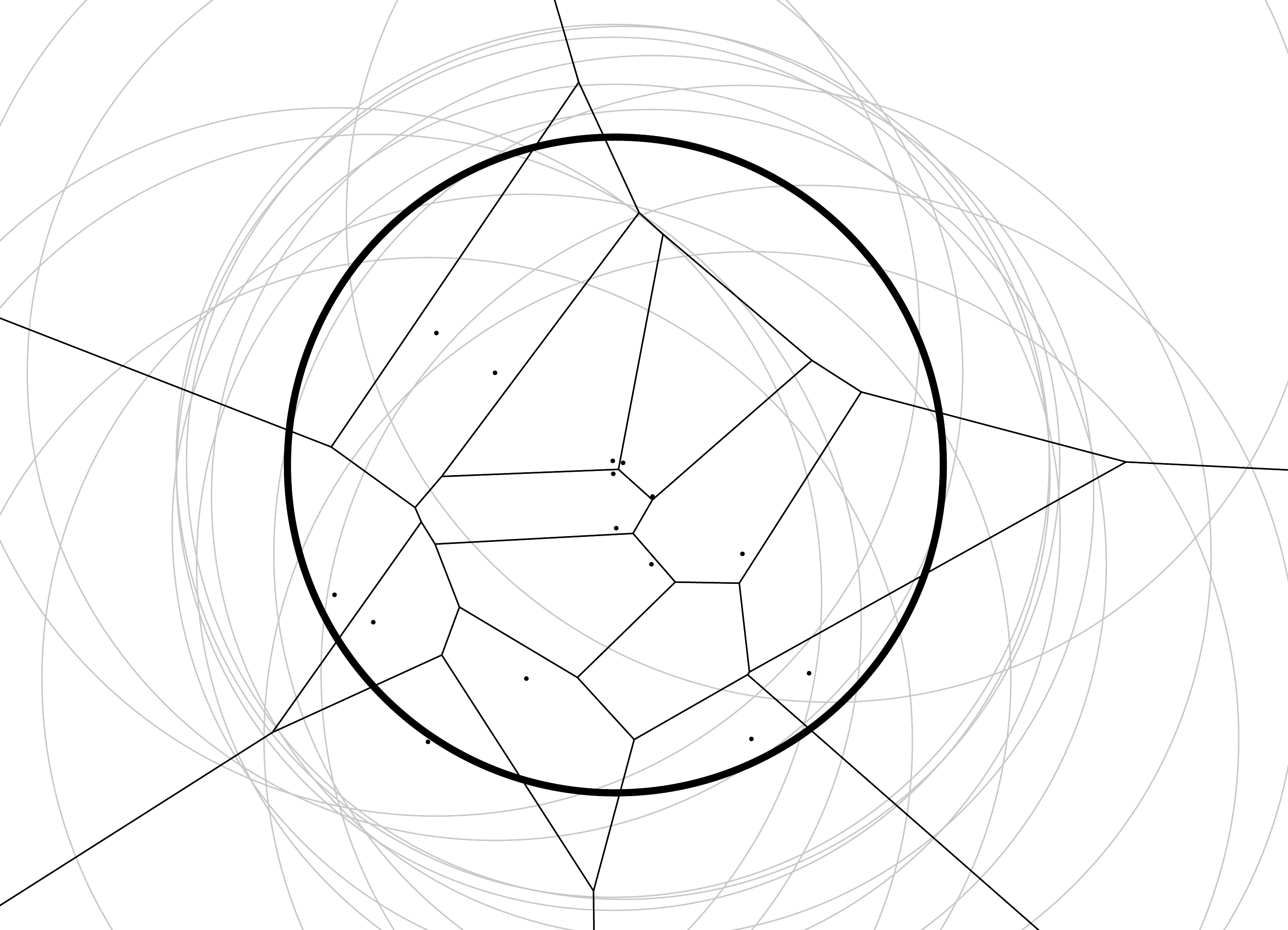} 
\\
(a) & (b) & (c)
\end{tabular}
\caption{The hyperbolic Voronoi diagram in conformal Poincar\'e disk (a) is obtained by a radial scaling transformation of the HVD in non-conformal Klein disk (b) that is itself built as an equivalently clipped power diagram (c). Observed that some bounded cells of the power diagram are cut by the boundary cutting circle.
}%
\label{fig:clipping}%
\end{figure}

Note that  clipping may destroy bounded cells of the affine diagram as illustrated in Figure~\ref{fig:clipping}.
Thus a remaining open question is to report an optimal output-sensitive construction of the $k$-order HVDs.

A video illustrating the hyperbolic Voronoi diagrams using the five common models of hyperbolic geometry is available online~\cite{HVDvideo}. 
 
\nocite{*}

\bibliographystyle{plain}

% that's all folks
\end{document}